\documentclass[12pt, preprint]{aastex}

\usepackage{graphicx}
\usepackage{natbib}
\bibpunct{(}{)}{;}{a}{}{,}
\shorttitle{Rb, Cd, and Ge in Planetary Nebulae}
\shortauthors{Sterling et al.}

\begin{document}

\title{Discovery of Rubidium, Cadmium, and Germanium Emission Lines in the Near-Infrared Spectra of Planetary Nebulae\footnote{This paper includes data taken at The McDonald Observatory of The University of Texas at Austin.}}

\author{N.\ C.\ Sterling\altaffilmark{1}, Harriet L.\ Dinerstein\altaffilmark{2}, Kyle F.\ Kaplan\altaffilmark{2}, \& Manuel A.\ Bautista\altaffilmark{3}}

\altaffiltext{1}{Department of Physics, University of West Georgia, 1601 Maple Street, Carrollton, GA 30118; nsterlin@westga.edu}
\altaffiltext{2}{Department of Astronomy, University of Texas, 2515 Speedway, C1400, Austin, TX 78712-1205; harriet@astro.as.utexas.edu, kfkaplan@astro.as.utexas.edu}
\altaffiltext{3}{Department of Physics, Western Michigan University, Kalamazoo, MI 49008; manuel.bautista@wmich.edu}

\begin{abstract}

We identify $[$\ion{Rb}{4}$]$~1.5973 and $[$\ion{Cd}{4}$]$~1.7204 $\mu$m emission lines in high-resolution ($R=40,000$) near-infrared spectra of the planetary nebulae (PNe) NGC~7027 and IC~5117, obtained with the IGRINS spectrometer on the 2.7-m telescope at McDonald Observatory.  We also identify $[$\ion{Ge}{6}$]$~2.1930~$\mu$m in NGC~7027.  Alternate identifications for these features are ruled out based on the absence of other multiplet members and/or transitions with the same upper levels.  Ge, Rb, and Cd can be enriched in PNe by \emph{s}-process nucleosynthesis during the asymptotic giant branch (AGB) stage of evolution.  To determine ionic abundances, we calculate $[$\ion{Rb}{4}$]$ collision strengths and use approximations for those of $[$\ion{Cd}{4}$]$ and $[$\ion{Ge}{6}$]$.  Our identification of $[$\ion{Rb}{4}$]$ 1.5973~$\mu$m is supported by the agreement between Rb$^{3+}$/H$^+$ abundances found from this line and the 5759.55~\AA\ feature in NGC~7027.  Elemental Rb, Cd, and Ge abundances are derived with ionization corrections based on similarities in ionization potential ranges between the detected ions and O and Ne ionization states.  Our analysis indicates abundances 2--4 times solar for Rb and Cd in both nebulae.  Ge is subsolar in NGC~7027, but its abundance is uncertain due to the large and uncertain ionization correction.  The general consistency of the measured relative \emph{s}-process enrichments with predictions from models appropriate for these PNe (2.0--2.5~M$_{\odot}$, $[$Fe/H$]=-0.37$) demonstrates the potential of using PN compositions to test \emph{s}-process nucleosynthesis models.


\end{abstract}

\keywords{planetary nebulae: general---nucleosynthesis, abundances--- stars: AGB and post-AGB---atomic data--- infrared: general}

\section{INTRODUCTION} \label{intro}

Planetary nebulae (PNe) mark the transition of asymptotic giant branch (AGB) stars to white dwarfs in low- and intermediate-mass (1--8~M$_{\odot}$) star evolution.  Their compositions bear the signatures of nucleosynthesis and mixing events during the AGB.  In particular, He, C, N, and elements formed by slow neutron(\emph{n})-capture nucleosynthesis (the \emph{s}-process; atomic number $Z>30$) can be enriched in PNe \citep{busso99, herwig05, karakas14}.

The low cosmic abundances of \emph{n}-capture elements \citep{asplund09} rendered them elusive to detection in astrophysical nebulae until \citet{pb94} identified emission lines of several trans-iron species in the optical spectrum of the PN~NGC~7027.  However, the deep, high-resolution spectra needed to unambiguously identify optical \emph{n}-capture element lines \citep{sharpee07, otsuka11, garcia-rojas15} restrict such studies to relatively bright PNe.

The near-infrared (NIR) spectral region has proven more fruitful for studies of \emph{s}-process enrichments in a large number of PNe.  \citet{dinerstein01} identified two $K$~band emission lines in PNe as  $[$\ion{Kr}{3}$]$~2.1986 and $[$\ion{Se}{4}$]$~2.2864~$\mu$m.\footnote{In this paper we use vacuum wavelengths for NIR lines, and air wavelengths for optical lines.}  \citet{sterling08} detected these lines in 81 of 120 Galactic PNe, and derived Se and Kr abundances \citep[recently revised by][]{sterling15} in these objects.

In this Letter, we identify $[$\ion{Rb}{4}$]$~1.5973, $[$\ion{Cd}{4}$]$~1.7204, and $[$\ion{Ge}{6}$]$~2.1930~$\mu$m for the first time in PNe.  To our knowledge, the only previous detections of the Rb and Cd features were marginal: \citet{zhangkwok92} found a very weak line at 1.721~$\mu$m in IC~5117, while \citet{hora99} detected a feature at 1.597~$\mu$m at the 1.5-$\sigma$ level in NGC~7027.

The detection of Rb and Cd are of particular note.  Rb enrichments are sensitive to the \emph{s}-process neutron density and hence the neutron source ($\alpha$-captures onto $^{13}$C or $^{22}$Ne) and initial stellar mass \citep{gh06, vanraai12}.  Cd lies beyond the first \emph{s}-process enrichment peak, and its abundance relative to lighter \emph{n}-capture elements can be used to constrain the time-averaged neutron flux in the AGB progenitors of PNe \citep[e.g.,][]{busso01}.  These detections therefore provide new means to investigate the characteristics of \emph{s}-process nucleosynthesis in PN progenitor stars.

\section{OBSERVATIONS AND ANALYSIS} \label{obs}

The spectra were obtained with the Immersion GRating INfrared Spectrometer (IGRINS) on the 2.7-m Harlan J. Smith Telescope at McDonald Observatory. IGRINS provides complete simultaneous coverage of the $H$ and $K$ bands (1.45--2.45~$\mu$m) at spectral resolution $R \approx 40,000$ \citep{park14}. The fixed slit is $1'' \times 15 ''$ on the sky.  IC 5117 was observed on 2014 September 9 with an E-W slit orientation for a total integration time of 40 minutes.  The source was nodded along the slit to maximize observing efficiency.  NGC 7027 was observed on 2014 October 23, with the on-source integration time totaling 18 minutes. The slit was oriented along the major axis of the PN (PA 60$\degr$ E of N), with sky frames collected by nodding off the target for this more spatially extended PN.

The data were processed with the IGRINS Pipeline Package\footnote{\url{https://github.com/igrins/plp}} written by J.-J.\ Lee, after removal of cosmic rays.  We used ThAr arc lamps for wavelength calibration, with a small correction from OH sky lines.  Barycentric and systemic velocity shifts were removed using nebular lines with precisely known wavelengths.  A0V standards were observed at similar times and airmasses as the targets for relative flux calibrations and telluric corrections.  See Kaplan et al.\ (in preparation) for further details.

For NGC 7027, the flux-calibrated 2D data were transformed to position-velocity space with 1 km s$^{-1}$-wide pixels before extraction of line fluxes. The position-velocity diagrams for the trans-iron element lines are shown in Figure~\ref{lines}, along with the apertures used for flux extraction.  Line fluxes were measured by summing all pixels within each aperture and subtracting the median value of pixels within 100~km~s$^{-1}$ of the line but outside the aperture. Since IC 5117 is essentially a point source, its spectrum was optimally extracted in 1D, transformed to velocity space, and line fluxes found by summing the flux above background for pixels within $\pm 50$ km s$^{-1}$ of the line center.  

\section{IDENTIFICATION OF $[$\ion{Rb}{4}$]$~1.5973, $[$\ion{Cd}{4}$]$~1.7204, and $[$\ion{Ge}{6}$]$~2.1930~$\mu$m} 

\begin{figure}[ht!]
\epsscale{0.82}
\plotone{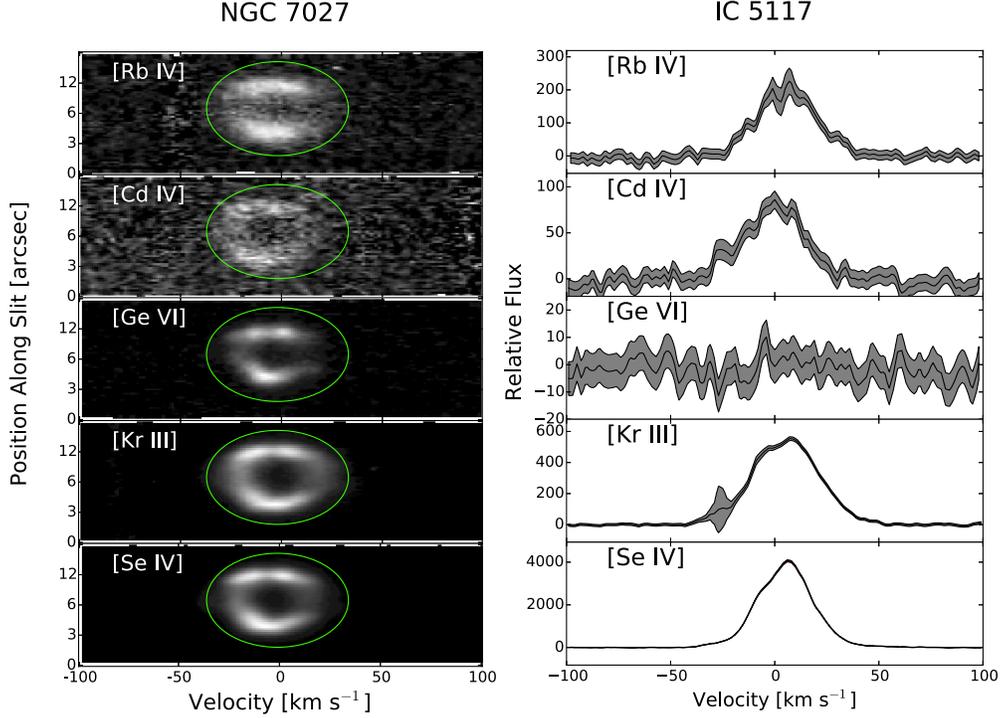}
\caption{\textit{Left panels:} 2D profiles of \emph{n}-capture element lines detected in NGC~7027, with apertures used for flux extraction.  \textit{Right panels:} 1D profiles of \emph{n}-capture transitions in IC~5117, including the non-detection of $[$\ion{Ge}{6}$]$.  Gray shading indicates 1-$\sigma$ uncertainties.}
\label{lines}
\end{figure}

We searched the IGRINS spectra of NGC~7027 and IC~5117 for \emph{n}-capture element emission lines with wavelengths computed from energy levels in the NIST Atomic Spectra Database\footnote{http://physics.nist.gov/asd} \citep{kramida14}.  We detected features at 1.5973 and 1.7203~$\mu$m in both objects and at 2.1935~$\mu$m in NGC~7027, which we associate with $[$\ion{Rb}{4}$]$~1.5973, $[$\ion{Cd}{4}$]$~1.7204, and $[$\ion{Ge}{6}$]$~2.1930~$\mu$m, respectively.  These features are shown along with $[$\ion{Kr}{3}$]$~2.1986 and $[$\ion{Se}{4}$]$~2.2864~$\mu$m in Figure~\ref{lines}.

We used the Atomic Line List v2.05b18\footnote{http://www.pa.uky.edu/$\sim$peter/newpage/} to search for alternate identifications within $\pm$10~\AA\ of the detected features.  We considered permitted features of ions in the first three rows of the Periodic Table, and forbidden transitions with upper-level excitation energies less than 10~eV. No molecular transitions match the wavelengths of these features, and telluric features can be readily distinguished from nebular lines at our spectral resolution.

For the possible alternate identifications for $[$\ion{Rb}{4}$]$~1.5973~$\mu$m, no additional multiplet members nor transitions with the same upper levels were seen.  In fact, no lines of the species considered (\ion{He}{1}, \ion{C}{2}, \ion{N}{1}, \ion{Ne}{2}, \ion{Ne}{4}, \ion{Ar}{2}, $[$\ion{Ti}{2}$]$, and $[$\ion{V}{1}$]$), are seen in the IGRINS spectra of NGC~7027 or IC~5117 with the exception of \ion{He}{1}.  Rb$^{3+}$ is created at an energy of 39.3~eV and destroyed at 52.2~eV.  This is comparable to the ionization potential (IP) range of O$^{2+}$ (35.1--54.9~eV), and therefore Rb$^{3+}$ should be the dominant Rb ion in moderate to high excitation PNe.  The agreement between Rb$^{3+}$/H$^+$ abundances derived from the 1.5973~$\mu$m line and the optical $\lambda$5759.55 transition (\S\ref{abund}) provides compelling support for this identification.

Cd is predicted to be highly enriched by \emph{s}-process nucleosynthesis models \citep[e.g.,][]{cristallo15}.  Cd$^{3+}$ has a $4d^9$ ground configuration with two levels separated by 5812.6~cm$^{-1}$ \citep{joshi77}, corresponding to a wavelength of 1.7204~$\mu$m.  Its IP range is 37.5--51.0~eV, similar to Rb$^{3+}$ and O$^{2+}$, and thus Cd$^{3+}$ should be the most abundant Cd ion in all but low-excitation PNe.  No multiplet members or lines from the same upper levels were seen for possible alternate identifications within $\pm$10~\AA\ of the 1.7203~$\mu$m feature.  Therefore, $[$\ion{Cd}{4}$]$~1.7204~$\mu$m is the most probable identification.

Finally, we detect a line at 2.1935~$\mu$m in NGC~7027.  While $[$\ion{Ge}{6}$]$~2.1930~$\mu$m does not perfectly match the observed wavelength, no multiplet members or lines with the same upper level are detected for other possible identifications.  Ge$^{5+}$ has a similar electronic structure to Cd$^{3+}$, with a $3d^9$~$^2$D ground term leading to a single forbidden line.  The IP range of 90.5--115.9~eV for Ge$^{5+}$ is similar to that of Ne$^{4+}$ (97.2--126.2~eV), which is detected in NGC~7027 \citep{zhang05} but not IC~5117 \citep{hyung01}.  It is therefore reasonable that $[$\ion{Ge}{6}$]$ is seen in the spectrum of NGC~7027 but not in IC~5117.

\section{ABUNDANCE ANALYSIS}

Table~\ref{lineinfo} lists fluxes for the \emph{n}-capture element lines detected in the IGRINS spectra of NGC~7027 and IC~5117.  Error bars to the line fluxes include $\sim$10\% uncertainties in the continuum placement in addition to statistical errors.  The $[$\ion{Kr}{3}$]$ and $[$\ion{Se}{4}$]$ fluxes are in excellent agreement with previous findings \citep{sterling08}.

\begin{deluxetable}{lccccc}
\tablecolumns{6}
\tablewidth{0pc} 
\tabletypesize{\footnotesize}
\tablecaption{Fluxes and Abundances of Neutron-Capture Elements in NGC~7027 and IC~5117} 
\tablehead{
\colhead{Line} & \colhead{Flux} & \colhead{Ionic} & \colhead{} & \colhead{Elemental} & \colhead{}\\
\colhead{Ratio} & \colhead{Ratio} & \colhead{Abund.\ X$^{i+}$/H$^+$} & \colhead{ICF\tablenotemark{a}} & \colhead{Abund.\ (X/H)} & \colhead{$[$X/H$]$}}
\startdata
\cline{1-6} 
\multicolumn{6}{c}{NGC~7027} \\
\cline{1-6}
$[$\ion{Rb}{4}$]$~1.5973 / \ion{H}{1}~1.5885 & (5.72$\pm$0.57)E--02 & (6.05$\pm$0.61)E--10 & 1.88 & (1.14$\pm$0.55)E--09 & 0.54$\pm$0.20 \\
$[$\ion{Rb}{4}$]$~5759.55 / H$\beta$\tablenotemark{b} & (1.68$\pm$0.43)E--04 & (7.70$\pm$1.78)E--10 & 1.88 & (1.45$\pm$0.70)E--09 & 0.64$\pm$0.20 \\
$[$\ion{Cd}{4}$]$~1.7204 / \ion{H}{1}~1.6811 & (5.14$\pm$0.54)E--03 & (5.74$\pm$0.60)E--11 & 1.88 & (1.08$\pm$0.81)E--10 & 0.32$\pm$0.30 \\
$[$\ion{Ge}{6}$]$~2.1930 / \ion{H}{1}~2.1655 & (5.61$\pm$0.56)E--03 & (2.77$\pm$0.28)E--10 & 6.52 & (1.81$\pm$2.58)E--09 & $-0.32\pm0.50$ \\
$[$\ion{Kr}{3}$]$~2.1980 / \ion{H}{1}~2.1655 & (2.95$\pm$0.30)E--02 & (2.28$\pm$0.23)E--09 & 5.09 & (1.16$\pm$0.26)E--08 & 0.81$\pm$0.10 \\
$[$\ion{Se}{4}$]$~2.2858 / \ion{H}{1}~2.1655 & (7.87$\pm$0.79)E--02 & (1.32$\pm$0.13)E--09 & 3.57 & (4.72$\pm$1.06)E--09 & 0.34$\pm$0.10 \\
\cline{1-6}
\multicolumn{6}{c}{IC~5117} \\
\cline{1-6}
$[$\ion{Rb}{4}$]$~1.5973 / \ion{H}{1}~1.5885 & (5.40$\pm$0.64)E--02 & (7.61$\pm$0.84)E--10 & 1.25 & (9.52$\pm$4.54)E--10 & 0.46$\pm$0.20 \\
$[$\ion{Cd}{4}$]$~1.7204 / \ion{H}{1}~1.6811 & (1.12$\pm$0.12)E--02 & (1.37$\pm$0.15)E--10 & 1.25 & (1.71$\pm$1.29)E--10 & 0.52$\pm$0.30 \\
$[$\ion{Kr}{3}$]$~2.1980 / \ion{H}{1}~2.1655 & (2.21$\pm$0.23)E--02 & (1.89$\pm$0.20)E--09 & 3.78 & (7.13$\pm$1.61)E--09 & 0.60$\pm$0.10 \\
$[$\ion{Se}{4}$]$~2.2858 / \ion{H}{1}~2.1655 & (1.31$\pm$0.13)E--01 & (2.39$\pm$0.24)E--09 & 1.73 & (4.14$\pm$0.92)E--09 & 0.28$\pm$0.10 \\
\tableline
\enddata
\label{lineinfo}
\tablecomments{Neutron-capture element line fluxes relative to nearby \ion{H}{1} lines are given, as are ionic abundances, ionization correction factors, and elemental abundances relative to solar \citep{asplund09}.  Vacuum wavelengths in $\mu$m are used for NIR lines, and air wavelengths in \AA\ for optical transitions.}
\tablenotetext{a}{We use ICF(Rb, Cd)~=~O/O$^{2+}$, based on the similar IP ranges of Rb$^{3+}$, Cd$^{3+}$, and O$^{2+}$, and ICF(Ge)~=~Ne/Ne$^{4+}$ for the same reason.  The O$^{2+}$, Ne$^{4+}$, Kr$^{2+}$, and Se$^{3+}$ ionic fractions are taken from Cloudy models of these PNe \citep{sterling15}.}
\tablenotetext{b}{From \citet{sharpee07}, with the average of the \ion{He}{2}~$\lambda \lambda$5757.02, 5762.64 intensities subtracted from the observed intensity to account for contamination from \ion{He}{2}~$\lambda$5759.74.}
\end{deluxetable}
\clearpage

\subsection{Ionic and Elemental Abundances} \label{abund}

We compute ionic abundances (Table~\ref{lineinfo}) relative to H$^+$ using nearby \ion{H}{1} lines, assuming an electron temperature $T_{\rm e}=12600\pm 500$~K and density $n_{\rm e}=52300$~cm$^{-3}$ for NGC~7027 \citep{zhang05}, and $T_{\rm e}=11800\pm 300$~K and $n_{\rm e}=89000$~cm$^{-3}$ for IC~5117 \citep{hyung01}.  Error bars for the ionic abundances include uncertainties in $n_{\rm e}$ (assumed to be 20\%), $T_{\rm e}$, and line fluxes; the latter dominates the uncertainties due to the weak dependence of these transitions on temperature and density.  We utilize the transition probabilities of \citet{biemont86b} for $[$\ion{Kr}{3}$]$, \citet{biemont87} for $[$\ion{Se}{4}$]$ and $[$\ion{Ge}{6}$]$, and the collision strengths of \citep{schoning97} and K.\ Butler (2007, private communication) for $[$\ion{Kr}{3}$]$ and $[$\ion{Se}{4}$]$ respectively.  We have computed effective collision strengths and transition probabilities for $[$\ion{Rb}{4}$]$, as described in the Appendix.  For $[$\ion{Ge}{6}$]$, we approximate the collision strength by scaling that of the isoelectronic ion $[$\ion{Se}{4}$]$ according to the ratio of effective nuclear charges, following the procedure of \citet{sharpee07} for lines of Br ions.  We adopt the $[$\ion{Se}{4}$]$ collision strength and $[$\ion{Ge}{6}$]$ transition probability for $[$\ion{Cd}{4}$]$.  

Our $[$\ion{Rb}{4}$]$ collision strengths can be tested by comparing the strength of the 1.5973~$\mu$m line to the 5759.55~\AA\ transition.  The optical line was detected in NGC~7027 \citep{sharpee07}, but not IC~5117 \citep{hyung01}.  From the 1.5973~$\mu$m intensity, we predict the intensity of $[$\ion{Rb}{4}$]$~5759.55~\AA\ in NGC~7027 to be $(1.32\pm 0.13) \times 10^{-4}$ relative to H$\beta$, using the theoretical \ion{H}{1}~Br15 to H$\beta$ ratio \citep{storey95} at $T_{\rm e}=12600$~K and $n_{\rm e}=52300$~cm$^{-3}$.  We estimate the contribution of \ion{He}{2}~$\lambda$5759.74 to the optical feature's measured flux by averaging the fluxes of the adjacent \ion{He}{2} $\lambda \lambda$5762.64 and 5757.02 lines.  After correcting for the blended \ion{He}{2} line, the residual $[$\ion{Rb}{4}$]$~5759.55~\AA\ intensity is $(1.68\pm0.43)\times 10^{-4}$.  The predicted and observed intensities agree within the uncertainties, supporting the accuracy of our collision strength calculations and the identification of the 1.5973~$\mu$m feature as $[$\ion{Rb}{4}$]$.

To derive elemental abundances (Table~\ref{lineinfo}), unobserved ions must be accounted for via ``ionization correction factors'' (ICFs).  Reliable ICF prescriptions require accurate photoionization cross sections and recombination rate coefficients, but these data are currently unknown for Rb, Cd, and Ge ions.  We therefore adopt ICFs based on similarities in IP ranges: ICF(Rb, Cd)~=~Rb/Rb$^{3+}$~=~Cd/Cd$^{3+}$~~=~O/O$^{2+}$ and ICF(Ge)~=~Ge/Ge$^{5+}$~=~Ne/Ne$^{4+}$.  In the absence of reliable atomic data, it is difficult to quantify the effects of uncertainties in the adopted ICFs and the collision strengths for $[$\ion{Ge}{6}$]$ and $[$\ion{Cd}{4}$]$ on the abundance determinations.  Based on the uncertainties in the Se and Kr abundances (0.1~dex), we roughly estimate uncertainties of 0.2~dex for Rb, 0.3~dex for Cd, and 0.5~dex for Ge (whose ICF is large and very uncertain).

The O$^{2+}$, Ne$^{4+}$, Kr$^{2+}$, and Se$^{3+}$ ionic fractions required to estimate ICFs were extracted from Cloudy models of NGC~7027 and IC~5117 \citep{sterling15}.  Our derived Kr and Se abundances agree with the empirical values found by \citet{sterling15} to within 15\%.  The Rb and Cd abundances are larger than solar \citep{asplund09} in both PNe by amounts (0.3--0.6~dex) similar to those of Se and Kr.  The nominally subsolar Ge abundance in NGC~7027 is not a firm result due to the large and uncertain ICF caused by the fact that Ge$^{5+}$ is a minority species.  Depletion into dust may also be a factor for this moderately refractory element \citep{lodders03}.

Depletion is unlikely to affect Cd due to its low condensation temperature \citep[652~K;][]{lodders03} and mild depletion in the diffuse interstellar medium \citep{sofia99}.  The situation is less clear for Rb, which has a condensation temperature of 800~K.  Other alkali elements such as Na and K are depleted by factors of 2--4 in PNe \citep{pottasch09, garcia-rojas15}, including NGC~7027 \citep{bernard-salas01}.  But Na and K have higher condensation temperatures (950--1000~K) than Rb.  We conclude that Rb may be depleted, but probably by less than a factor of two (the typical depletion of Na in PNe).

\subsection{Comparison With Models} \label{modelcomp}

Including optical Xe detections \citep{sharpee07, hyung01}, 5--6 \emph{n}-capture elements have been detected in NGC~7027 and IC~5117.  This allows for a systematic comparison of the observed enrichment patterns with theoretical predictions.  We compare our results with models from the FRUITY database \citep{cristallo11, cristallo15}\footnote{fruity.oa-teramo.inaf.it}, which samples initial stellar masses in the range 1.3--6.0~M$_{\odot}$ and metallicities $Z=0.001$--0.020.  We select the models with $Z=0.006$, which corresponds to $[$Fe/H$]=-0.37$ for $Z_{\odot}=0.014$.  This is consistent with the $[$Zn/H$]$ values -- a proxy for $[$Fe/H$]$ \citep{dinerstein_geballe01} -- for NGC~7027 and IC~5117: $[$Zn/H$]=-0.44\pm 0.08$ and $-0.31\pm 0.10$, respectively \citep{smith14}.  

\begin{figure}[t!]
\epsscale{0.85}
\plotone{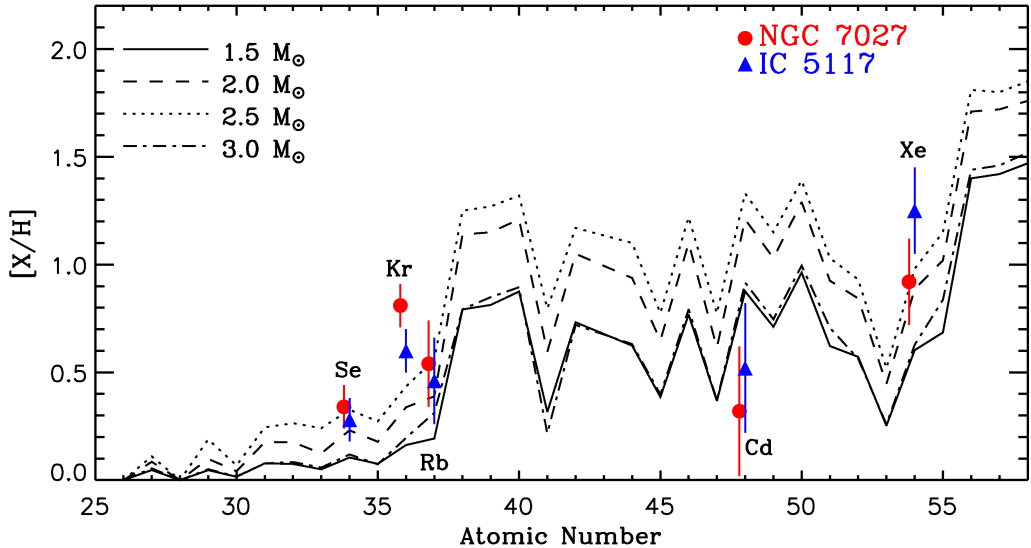}
\caption{Comparison of empirical and theoretical \emph{s}-process enrichments of \emph{n}-capture elements in NGC~7027 (red circles) and IC~5117 (blue triangles).  Data points for NGC~7027 are offset horizontally so that their error bars can be distinguished from those of IC~5117.  Theoretical predictions $[$X/Fe$]$ are from FRUITY models \citep{cristallo11, cristallo15} with metallicity $Z=0.006$ ($[$Fe/H$]=-0.37$).}
\label{thy_comp}
\end{figure}

In Figure~\ref{thy_comp} we plot predicted FRUITY final envelope abundances \citep{cristallo15} for AGB stars of initial mass 1.5--3.0~M$_{\odot}$ and metallicity Z=0.006.  The overall \emph{n}-capture enrichments first increase with mass, peaking at 2.0--2.5~M$_{\odot}$, then decrease.  We also plot the derived \emph{n}-capture element abundances for NGC~7027 and IC~5117 (Ge is not shown due to its uncertain abundance).  We adopt $[$Xe/Ar$]=0.92$ for NGC~7027 \citep{sharpee07}.  For IC~5117, we use the transition probabilities of \citet{biemont95} and collision strengths of \citet{sb98} to compute Xe$^{3+}$/H$^+=1.03\times10^{-9}$ from the $[$\ion{Xe}{4}$]$~$\lambda \lambda$5709.21, 7535.49 lines detected by \citet{hyung01}.  Assuming that Xe$^{3+}$/Xe~$\approx$~Ar$^{3+}$/Ar, we find Xe/H~=~$3.12\times 10^{-9}$ or $[$Xe/H$]=1.25$, with an estimated uncertainty of 0.2~dex due to the approximate ICF.

The theoretical curves represent $[$X/Fe$]$, while the measured abundances are given as $[$X/H$]$.  To convert the latter to $[$X/Fe$]$ one would have to add 0.37 to $[$X/H$]$ for each data point, which would raise the measured abundances above the predicted values.  However, the absolute abundances at the end of the thermally-pulsing AGB vary substantially among different sets of models for the same mass and metallicity, due to different treatments of mass loss, mixing at convective interfaces, and the number of thermal pulses before envelope ejection \citep{karakas14}.  Therefore the measured abundances should be compared with the relative enrichments in Figure~\ref{thy_comp}.

The relative Se, Rb, and Xe enrichments are consistent with an initial mass of 2.0--2.5~M$_{\odot}$ for NGC~7027, in agreement with estimates of $\sim2.5$M$_{\odot}$ based on its central star and nebular masses \citep{zijlstra08, santander-garcia12}.  The Se and Rb abundances of IC~5117 suggest a lower initial mass, as expected based on its central star mass \citep{hyung01}.  Interestingly, the measured Kr abundances are higher than predicted for both objects, as is the Xe abundance for IC~5117.  The reason for these discrepancies is unclear, especially since the Kr abundances are well-constrained given the detection of multiple Kr ions in each object \citep[e.g.,][]{sterling15}.  The derived abundance for Cd is not in agreement with theoretical predictions, most likely due to inaccuracies in our adopted $[$\ion{Cd}{4}$]$ collision strength and ICF.

Ratios between enrichments of elements in the first (``light,'' or ls) and second (``heavy,'' or hs) \emph{s}-process peaks are less sensitive to uncertainties in the physics of AGB models than absolute abundances.  However, indices such as $[$hs/ls$]$ used for studying \emph{s}-process nucleosynthesis in stars are based on elements that cannot be measured reliably in nebulae, for reasons including depletion into dust.  In nebulae, the ls peak is represented by Ge through Rb, while only Cd and Xe \citep[and perhaps Ba;][]{pb94} have been detected among the elements beyond this peak.  As we extend the sample of observed nebulae to include a wider range of initial stellar mass and metallicity, we plan to test the dependence of \emph{s}-process enrichments on these parameters, and compare measured enrichments with different sets of nucleosynthesis models.

\section{SUMMARY}

We identify NIR $[$\ion{Rb}{4}$]$, $[$\ion{Cd}{4}$]$, and $[$\ion{Ge}{6}$]$ emission lines in the spectra of the PNe NGC~7027 and IC~5117.  The identification of Rb is important due to the sensitivity of its enrichment to the \emph{s}-process neutron density and hence progenitor mass.  Since Cd lies beyond the first (``light'') \emph{s}-process peak, its abundance relative to light \emph{n}-capture elements is sensitive to the time-averaged neutron flux experienced by Fe nuclei during the \emph{s}-process.

We derive ionic abundances using newly-computed effective collision strengths for $[$\ion{Rb}{4}$]$ (see Appendix) and estimated collision strengths for $[$\ion{Cd}{4}$]$ and $[$\ion{Ge}{6}$]$.  The identification of $[$\ion{Rb}{4}$]$~1.5973~$\mu$m is strengthened by the agreement between the Rb$^{3+}$/H$^+$ abundance found from this line and an optical $[$\ion{Rb}{4}$]$ feature at 5759.55~\AA.  We derive elemental abundances for Rb, Cd, and Ge based on similarities between the IP ranges of the detected ions and O$^{2+}$ and Ne$^{4+}$.  The Rb and Cd abundances are factors of 2--4 greater than solar, comparable to Kr and Se.  Ge is subsolar by about a factor of two in NGC~7027, due to the uncertain ICF and/or depletion into dust.  The abundances relative to H are approximately a factor of two larger than predicted by FRUITY models \citep{cristallo15}, but the relative enrichments are in reasonable agreement.  The Cd abundances nominally fall below the theoretical curves, but are highly uncertain due to the current lack of atomic data.  Kr and Xe appear to be slightly more enriched than predicted by this set of models, an effect that merits further investigation.  These identifications provide new tools for studying \emph{s}-process enrichments in PNe, and demonstrate the potential for using the growing number of \emph{n}-capture elements detected in PNe to constrain and improve theoretical models of AGB evolution and nucleosynthesis.

\acknowledgements

NCS and MAB acknowledge support from the NSF through award AST-1412928, and HLD from AST-0708425.  This work used the Immersion Grating Infrared Spectrograph (IGRINS) developed by the University of Texas at Austin and the Korea Astronomy and Space Science Institute (KASI) with the financial support of NSF grant AST-1229522, the University of Texas at Austin, and the Korean GMT Project of KASI.  This work has made use of NASA's Astrophysics Data System, the Atomic Line List v2.05B18, and the FRUITY Database of nucleosynthetic yields from AGB stars.

\appendix \label{appx}

\section{$[$\ion{Rb}{4}$]$ EFFECTIVE COLLISION STRENGTHS}

We compute radiative rates (A-values) for dipole forbidden transitions among the $^3$P, $^1$D, and $^1$S levels of the $4s^24p^4$ configuration of Rb$^{3+}$ using the code AUTOSTRUCTURE \citep{badnell86, badnell11}.  We allow for orbital relaxation and configuration mixing by including the configurations $4s4p^5$, $4s^24p^34d$, $4s4p^44d$, $4s^24p^34f$, $4s^24p^35s$, $4s^24p^35p$, and $4s^24p^35d$.  This expansion yields level energies in good agreement with experimental values from NIST \citep{kramida14}, but gives A-values that are a factor of two lower than those of \citet{biemont86b}.  To produce more accurate A-values, it was necessary to include the configurations $3d^94s^24p^5$, $3d^94s^24p^44d$, and $3d^94s^24p^45s$.

The orbital wave functions were optimized by minimizing the energies of the eight lowest LS terms, and were fine-tuned by means of term energy corrections.  Energy levels were then shifted to match experimental values, and A-values (Table~\ref{rbiv_data}) were computed.  The predicted energy levels agree with experimental values to within ∼10\% up to about 0.55 Ryd, and better for higher energy terms.

Collision strengths for Rb$^{3+}$ were computed with the BPRM+ICFT method, using a suite of parallel Breit-Pauli R-matrix programs \citep{mitnik01, mitnik03, badnell04}.  We used the wave functions from the AUTOSTRUCTURE calculation, retaining configuration interaction from all 10 configurations and the lowest 23 LS terms from the $4s^24p^4$, $4s4p^5$ and $4s^24p^34d$ configurations. The calculations explicitly include partial waves from states with $L\leq 9$ and multiplicities 2, 4, and 6.  The final collision strengths (Table~\ref{rbiv_data}) were produced with an energy resolution of $9\times 10^{-4}$~Ryd up to an energy three times that of the highest threshold.  Full details of these calculations will be given in a forthcoming paper.

\begin{deluxetable}{ccc|ccccc}
\tablecolumns{8}
\tablewidth{0pc} 
\tablenum{A1}
\tabletypesize{\scriptsize}
\tablecaption{Transition Probabilities and Effective Collision Strengths for the $[$\ion{Rb}{4}$]$ $4s^24p^4$ Ground Configuration} 
\tablehead{
\colhead{Lower} & \colhead{Upper} & \multicolumn{1}{c|}{$A_{ul}$} & \multicolumn{5}{c}{Effective Collision Strengths}\\ \cline{4-8}
\colhead{Level} & \colhead{Level} & \multicolumn{1}{c|}{(s$^{-1}$)} & \colhead{5000~K} & \colhead{7500~K} & \colhead{10000~K} & \colhead{15000~K} & \colhead{20000~K}}
\startdata
$^3$P$_2$ & $^3$P$_1$ & 5.374 & 1.908 & 2.105 & 2.314 & 2.704 & 3.025 \\
$^3$P$_2$ & $^3$P$_0$ & 0.000 & 0.6836 & 0.7090 & 0.7521 & 0.8487 & 0.9373 \\
$^3$P$_2$ & $^1$D$_2$ & 5.324 & 3.508 & 3.417 & 3.343 & 3.368 & 3.481 \\
$^3$P$_2$ & $^1$S$_0$ & 0.000 & 0.6057 & 0.6709 & 0.6982 & 0.7024 & 0.6856 \\
$^3$P$_1$ & $^3$P$_0$ & 1.720E--2 & 0.6613 & 0.7092 & 0.7674 & 0.8692 & 0.9443 \\
$^3$P$_1$ & $^1$D$_2$ & 0.4775 & 2.062 & 2.000 & 1.977 & 2.006 & 2.071 \\
$^3$P$_1$ & $^1$S$_0$ & 32.51 & 0.3921 & 0.4287 & 0.4468 & 0.4542 & 0.4471 \\
$^3$P$_0$ & $^1$D$_2$ & 9.635E--5 & 0.7158 & 0.7034 & 0.6978 & 0.7037 & 0.7181 \\
$^3$P$_0$ & $^1$S$_0$ & 0.000 & 0.1499 & 0.1669 & 0.1796 & 0.1938 & 0.2000 \\
$^1$D$_2$ & $^1$S$_0$ & 5.907 & 1.293 & 1.322 & 1.346 & 1.403 & 1.470 \\
\tableline
\enddata
\label{rbiv_data}
\end{deluxetable}
\clearpage

\bibliographystyle{apj}


\end{document}